\documentclass[10pt,journal,compsoc]{IEEEtran}

\ifCLASSOPTIONcompsoc
  \usepackage[nocompress]{cite}
\else
  \usepackage{cite}
\fi

\usepackage[colorlinks,urlcolor=blue,linkcolor=blue,citecolor=blue]{hyperref}
\usepackage[symbol]{footmisc} 

\usepackage{todonotes}

\begin{document}

\title{Giving RSEs a Larger Stage through the Better Scientific Software Fellowship
\thanks{This manuscript has been authored by UT-Battelle, LLC, under contract DE-AC05-00OR22725 with the US Department of Energy (DOE). The publisher acknowledges the US government license to provide public access under the DOE Public Access Plan (\url{https://energy.gov/downloads/doe-public-access-plan}).}
}

\markboth{Computing in Science \& Engineering }{Special Issue}

\author{\IEEEauthorblockN{William F. Godoy\IEEEauthorrefmark{1}, 
Ritu Arora\IEEEauthorrefmark{2},
Keith Beattie\IEEEauthorrefmark{3},
David E. Bernholdt\IEEEauthorrefmark{1},
Sarah E. Bratt\IEEEauthorrefmark{4},
Daniel S. Katz\IEEEauthorrefmark{5},
Ignacio Laguna\IEEEauthorrefmark{6},
Amiya K. Maji\IEEEauthorrefmark{7},
Addi Malviya Thakur\IEEEauthorrefmark{1},
Rafael M. Mudafort\IEEEauthorrefmark{8},
Nitin Sukhija\IEEEauthorrefmark{9},
Damian Rouson\IEEEauthorrefmark{3},
Cindy Rubio-González\IEEEauthorrefmark{10},
Karan Vahi\IEEEauthorrefmark{11}
}

\IEEEauthorblockA{\IEEEauthorrefmark{1}Oak Ridge National Laboratory}
\IEEEauthorblockA{\IEEEauthorrefmark{2}The University of Texas at San Antonio}
\IEEEauthorblockA{\IEEEauthorrefmark{3}Lawrence Berkeley National Laboratory}
\IEEEauthorblockA{\IEEEauthorrefmark{4}University of Arizona}
\IEEEauthorblockA{\IEEEauthorrefmark{5}University of Illinois at Urbana-Champaign}
\IEEEauthorblockA{\IEEEauthorrefmark{6}Lawrence Livermore National Laboratory}
\IEEEauthorblockA{\IEEEauthorrefmark{7}Purdue University}
\IEEEauthorblockA{\IEEEauthorrefmark{8}National Renewable Energy Laboratory}
\IEEEauthorblockA{\IEEEauthorrefmark{9}Slippery Rock University of Pennsylvania}
\IEEEauthorblockA{\IEEEauthorrefmark{10}University of California, Davis}
\IEEEauthorblockA{\IEEEauthorrefmark{11}University of Southern California}
}

\IEEEtitleabstractindextext{%
\begin{abstract}
The Better Scientific Software Fellowship (BSSwF) was launched in 2018 to foster and promote practices, processes, and tools to improve developer productivity and software sustainability of scientific codes. BSSwF's vision is to grow the community with practitioners, leaders, mentors, and consultants to increase the visibility of scientific software production and sustainability. Over the last five years, many fellowship recipients and honorable mentions have identified as research software engineers (RSEs). This paper provides case studies from several of the program's participants to illustrate some of the diverse ways BSSwF has benefited both the RSE and scientific communities. In an environment where the contributions of RSEs are too often undervalued, we believe that programs such as BSSwF can be a valuable means to recognize and encourage community members to step outside of their regular commitments and expand on their work, collaborations and ideas for a larger audience.
\end{abstract}

}

\maketitle

\section{Introduction}
In 2017, the U.S.~Exascale Computing Project (ECP)\footnote{\url{https://www.exascaleproject.org}} funded the IDEAS Productivity project\footnote{The original project title is Interoperable Design of Extreme-Scale Application Software.} to support developer productivity, software sustainability, and reliability for the ECP's extensive scientific software and application portfolio, including more than 100 software packages and 24 applications.  The ECP encompasses roughly 1,000 people across the Department of Energy (DOE), national laboratories, academia, and industry, including both research software engineers (RSEs) and researchers from a broad range of scientific domains as well as computer science and applied mathematics.  

Given the scale and breadth of the challenge, the IDEAS project includes a wide array of ``outreach" activities to complement the relatively modest level of direct engagement that the IDEAS team can sustain with ECP software teams.~\cite{Heroux:2020:ASP}  IDEAS outreach supports the ECP project in general and the larger community of scientific software developers -- which constitute the workforce pool that ECP and the national laboratories draw upon -- as well as consumers and developers of dependencies of ECP software.  The three basic thrusts of the IDEAS outreach effort include (1) providing an online portal for the exchange of information and resource on scientific software development\footnote{\url{https://bssw.io}}, (2) offering seminars\footnote{\url{https://ideas-productivity.org/events/hpc-best-practices-webinars}} and training\footnote{\url{https://bssw-tutorial.github.io/}} to help developers improve their skills, and (3) raising the level of awareness and discussion of software-related issues in the community.  The third thrust involves substantial event organization -- e.g., creating opportunities within technical meetings for people to discuss their experiences with software development\footnote{\url{https://betterscientificsoftware.github.io/swe-cse-bof/}} as opposed to the typical focus on the scientific advances.  But we also look for ways to support members of the community who focus on the software side (who often identify as RSEs) and to amplify voices in the community.

Many aspects of the IDEAS project were inspired by the work of the Software Sustainability Institute (SSI)\footnote{\url{https://software.ac.uk}} in the United Kingdom.  The SSI has been sponsoring a Fellowship Program since 2012, with the goal to ``improve and promote good computational practice across all research disciplines and support those who are doing this important  work.''\footnote{\url{https://software.ac.uk/programmes-and-events/fellowship-programme}} In the fall of 2017, the IDEAS leadership began to formulate a fellowship program of their own with the idea of bringing new ideas and new voices to the DOE/ECP scientific software community, and in 2018 awards were made for the first class of the Better Scientific Software (BSSw) Fellowship.

The BSSw Fellowship (BSSwF) is somewhat different in structure and goals from the SSI Fellowship. The BSSwF program ``fosters and promotes practices, processes, and tools to improve developer productivity and software sustainability of scientific codes.''\footnote{\url{https://bssw.io/pages/bssw-fellowship-program}} Recognizing differences in how researchers are supported between the UK and the US, the BSSwF supports fewer fellows with larger stipends. This approach allows the fellows to pursue activities that are independent of leveraging other funding sources that may be available to them as a lesser amount might require to achieve meaningful results.  Each class of the BSSwF includes both fellows and honorable mentions.  Honorable mentions do not receive funding but are included in many of the networking opportunities arranged by the IDEAS project to help both fellows and honorable mentions get more exposure to the DOE, ECP, and the National Science Foundation (NSF) software communities, and identify opportunities for collaboration. The BSSwF encourages applications from people at all career stages. Awards have been made to graduate students, early career, mid-career and senior-level professionals. Likewise, awards have been made to individuals in academia, national laboratories, and small businesses.  Honorable mentions are eligible to reapply, but fellows can receive the award only once. Due to the nature of fellowship funding, the primary limitation on applicants is that they must be associated with a U.S.-based institution capable of receiving federal funds.

The size of the fellowship classes has varied over the years based primarily on the funding available.  We started with four fellows and four honorable mentions in each of the 2018 and 2019 classes, while in 2020 we supported three sets. In 2021, NSF joined the DOE as a co-sponsor of the fellowship, allowing a class of four and then six sets in 2022.

The present paper is co-authored by eight fellows, five honorable mentions, and one member of the fellowship's Executive Committee to illustrate the impact of the BSSwF to develop, connect and grow a community that recognizes efforts toward better scientific software. Each recipient provides an introduction and personal account on their experiences around BSSwF in Section~\ref{sec:Experiences}. Section~\ref{sec:Synergies} provides a brief account of the synergies and some of the lessons learned resulting from BSSwF activities. Finally, Section~\ref{sec:Conclusions} outlines our shared view advocating for more initiatives like BSSwF to elevate and promote RSE stewardship efforts on mission-critical software.

\section{Experiences}
\label{sec:Experiences}
The goal of this section is to showcase the recipients' personal experiences. Each account illustrates the diverse drivers, interests, and the efforts of the people from different backgrounds and career stages contributing towards building a better scientific software community that elevates the work of RSEs.

\subsection{Fellows}

\paragraph*{\textbf{Daniel S. Katz}} is Chief Scientist at the National Center for Supercomputing Applications (NCSA) and Research Associate Professor in Computer Science, Electrical and Computer Engineering, and the School of Information Sciences (iSchool) at the University of Illinois Urbana-Champaign. Dan received the B.S., M.S., and Ph.D degrees in Electrical Engineering from Northwestern University in 1988, 1990, and 1994, respectively.
His interest is in the development and use of advanced cyberinfrastructure to solve challenging problems at multiple scales, including applications, algorithms, fault tolerance, and programming in parallel and distributed computing, as well as policy issues such as citation and credit mechanisms and practices associated with software and data, organization and community practices for collaboration, and career paths for computing researchers.
He is a co-founder and current Associate Editor-in-Chief of the Journal of Open Source Software (JOSS), co-founder of the US Research Software Engineer Association (US-RSE), and co-founder and steering committee chair of the Research Software Alliance (ReSA).

At the time of his fellowship (2018), Dan led the scientific software and applications division at NCSA, where he was concerned about career paths and recognition for RSEs. The BSSwF allowed Dan to fund and focus on activities that would recognize software as a valid scientific metric through the software citation principles initiative~\cite{SoftCitePrincples}. He co-led the FORCE11 Software Citation Implementation Working Group along side astronomers, physicists, and geoscientists to produce the principles and promote practices for software citation to recognize the RSE community behind these efforts. The more significant outcome from Dan's fellowship experience is that software citation is now becoming widely accepted in several major conference venues, while software developers and maintainers are encouraged to provide citable sources for their work. The FORCE11 Software Citation Implementation Working Group has continued this work (with Dan continuing to co-lead it,) as well as focusing on other groups, such as publishers, funders, and repositories and registries, to institutionalize citation practices among scientific software stakeholders.

\paragraph*{\textbf{Ignacio Laguna}} is a Computer Scientist at the Center for Applied Scientific Computing (CASC) at the Lawrence Livermore National Laboratory (LLNL), California. His main area of research is high-performance computing (HPC) programming models and systems. He is particularly interested in software correctness, program analysis, compilers, fault tolerance, and debugging. He received his PhD degree in Computer Engineering from Purdue University in 2012 and is a senior member of IEEE. Ignacio has won several awards, including the ACM/IEEE-CS George Michael Fellowship in 2014, the Hans Meuer Award for best research paper at the International Supercomputing Conference (ISC) 2019, and the Best Reproducibility Advancement Award at Supercomputing (SC) 2021. 

As part of his fellowship (2019), Ignacio organized a Tutorial on Floating-Point Analysis Tools for Scientific Software\footnote{\url{https://fpanalysistools.org}} at several venues including, the SC and PEARC conferences. The tutorial goal was to demo tools that allow programmers to get insight about how different aspects of floating-point arithmetic affect their code and how to fix potential bugs. It focused on three specific areas of floating-point analysis: 
(a) analysis of error bounds and reevaluation of expressions to reduce the error, (b) compiler-induced error due to optimizations, and (c) floating-point exceptions in accelerators (e.g. GPUs). Developing accurate and reliable scientific software is notoriously difficult and efforts such as Ignacio's enable stakeholders to ensure numerical reproducibility. The latter becomes an even bigger challenges with the existing heterogeneity of HPC systems. Thus posing a very difficult problem going from compiler optimizations to different precision arithmetic that can significantly affect
the final numerical results. The BSSwF helped him expand his network and made his tools more visible to HPC developers. As a fellow, Ignacio participated at several ECP annual meetings, which provided a venue to meet new people, socialize his technical ideas in the DOE HPC community, and come up with new research ideas to address important problems.

\paragraph*{\textbf{Damian Rouson}} is a Staff Scientist and the Group Leader for the Computer Languages and Systems Software (CLaSS) Group at Lawrence Berkeley National Laboratory (LBNL). He obtained his PhD in Mechanical Engineering at Stanford University in 1997. He is the technical lead for LBNL's LLVM Flang Fortran compiler testing project and is the lead developer of the Caffeine parallel runtime library. He is also the founder and President of Archaeologic Inc. and Sourcery Institute for which he leads research software engineering and software archaeology projects for academic, government, and industry clients. His research at LBNL explores ways to use machine learning to accelerate predictions of the regional impact of global climate change using the Intermediate Complexity Atmospheric Research model.

\begin{figure*}[ht]
\centering 
\includegraphics[clip, width=\linewidth,trim=0.5cm 3cm 2cm 3cm]{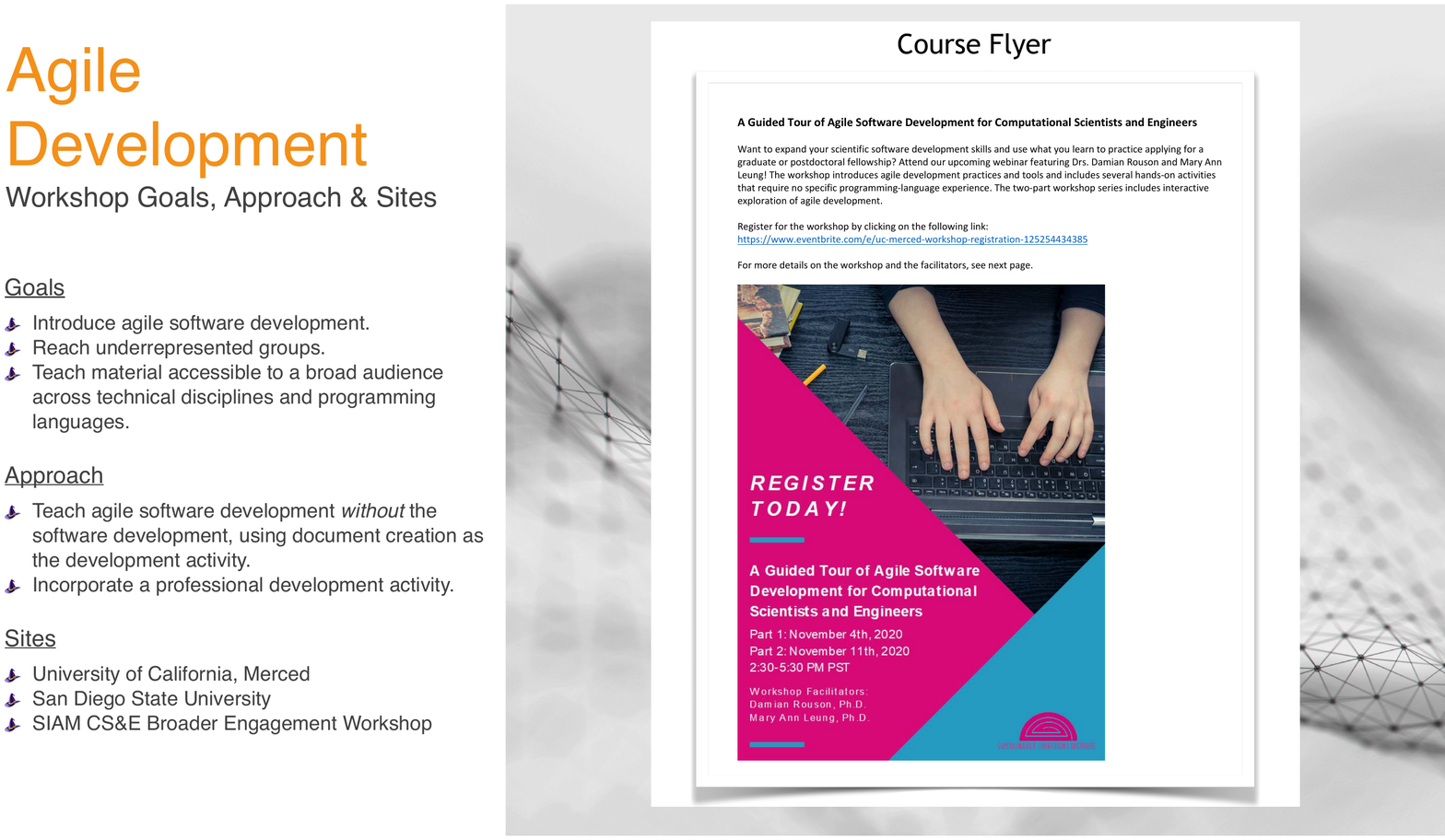}
\caption{Agile software development workshop slide from Damian Rouson's 2020 BSSwF project.}
\label{fig:agile}
\end{figure*}

At the time of his fellowship (2020), Damian worked part-time for Sustainable Horizons Institute (SHI), where he collaborated with SHI President Mary Ann Leung (also a BSSwF recipient), and Archaelogic engineer Brad Richardson on a series of workshops that introduced agile software development training. A novel aspect of these workshops was that there was no programming language prerequisite as advertised in the event flyer shown in Figure~\ref{fig:agile}. Instead the focus was on the collaborative process aspects of source control using git and GitHub workflows, including pull requests and continuous integration using a simple proxy project such as markdown document creation. Students were also exposed to concepts like pair programming, unit testing and test-driven development. One of the important goals of Damian's fellowship project was to attract students and scientists across a broad range of disciplines, programming backgrounds, career stages, and demographics with an emphasis on reaching out to groups that are underrepresented in computational science and engineering. The BSSwF enabled Damian to contribute to lowering the barriers to entry into projects related to scientific software. In this experience, he discovered that most of what one needs to learn to employ agile practices effectively does not require doing software development.  

\paragraph*{\textbf{Cindy Rubio-Gonz\'alez}} is an Associate Professor of Computer Science at the University of California, Davis. She received her PhD in Computer Science from the University of Wisconsin--Madison in 2012 followed by a postdoctoral appointment in the EECS Department at UC Berkeley. Her research spans the areas of Programming Languages, Software Engineering and HPC, with a focus on program analysis for automated bug finding and optimization. She is particularly interested in the reliability and performance of systems software and scientific applications. Cindy is a member of the ACM Special Interest Group on Programming Languages (SIGPLAN) Executive Committee and a recipient of several awards including the DOE Early Career Award, NSF CAREER Award, Facebook Testing and Verification Research Award, UC Davis Hellman Fellowship, and UC Davis CAMPOS Faculty Award.

During her fellowship (2020), Cindy focused on disseminating techniques for automated floating point mixed-precision tuning and software testing of numerical software. Mixed precision balances the trade-off between accuracy and cost of full numerical representations as used heavily by scientific applications. As part of her BSSw work, Cindy presented state-of-the-art efforts at various venues, and developed a graduate seminar that focuses on the use and evaluation of techniques to improve the reliability and performance of numerical software, with an emphasis on the reproducibility of experiments affected by numerical precision. The BSSw program provided Cindy with a unique opportunity to connect, share and learn from others who are also passionate about building better scientific software. Being part of the BSSw community has inspired Cindy to continue her work designing and developing tools that target a significant topic in scientific computing. The outcome of her work is to provide the community with tools that enable writing better, more efficient and reproducible scientific software.

\paragraph*{\textbf{Ritu Arora}} is the Assistant Vice President of Research Computing at the University of Texas at San Antonio. She received her PhD degree in Computer and Information Science from the University of Alabama at Birmingham in 2010. Prior to joining UTSA, Ritu worked at the University of Texas (UT) at Austin, where she was appointed as a Research Scientist at the Texas Advanced Computing Center (TACC). At TACC, she provided consultancy on effectively leveraging large-scale systems for activities ranging from big data management to video analytics, while teaching courses in scientific programming, parallel programming, and intelligent systems in geosciences.
Her areas of research and expertise include HPC, cloud computing, large-scale data management, and advanced software engineering. 

For Ritu's fellowship (2022), she will be developing training material on how to optimize input/output (I/O) in scientific applications targeting artificial intelligence/machine learning (AI/ML) workflows. Topics covered include: (a) optimizing I/O for data analysis and checkpointing in serial and parallel applications written in C, C++, Fortran, Python, and R, MPI, OpenMP, and CUDA; (b) optimizing I/O and checkpointing AI/ML models and workflows; and (c) techniques for leveraging the features in the underlying hardware and filesystems (e.g., Lustre) for optimizing applications’ I/O while being aware of portability issues. The material will be distributed through accessible online resources such as YouTube, GitHub-hosted repositories, LinkedIn and BSSw.io blog articles. The BSSwF program has been immensely valuable for Ritu for expanding her network of colleagues, especially in the DOE laboratories, and exploring opportunities of collaboration. The program has provided visibility to the cause that she is very passionate about which is ``optimizing IO" from serial and parallel applications in order to efficiently take advantage of the underlying HPC hardware as more data-driven AI/ML drive scientific discovery.

\paragraph*{\textbf{Nitin Sukhija}} is the Director of Center for Cybersecurity and Advanced Computing and an associate professor of computer science at Slippery Rock University of Pennsylvania. He graduated with a PhD in Computer Science from Mississippi State University in 2015 and has been involved in various research and management projects pertaining to the security and software challenges in industry and academia. His work aims to address the threats to software Confidentiality, Integrity, Availability in HPC environments. His research focuses on the integration of data analytics, vulnerability and risk assessment, autonomic computing, predictive control, and stochastic optimization techniques and models for achieving cyber-resilience. Sukhija chaired and co-chaired many conferences such as ACM and IEEE conferences while serving as an active committee member in several scientific computing and broader engagement venues. He is currently serving as the co-chair for the ACM SIGHPC Education Chapter workshop committee and has been active in the planning and participation in HPC Training Workshops series at the SC, ISC and other conferences since 2015.

During his fellowship (2022), Sukhija organized a tutorial on Best Practices and Tools for Secure Scientific Software Development at the SIAM Conference of Mathematical Data Science. This tutorial includes components for evaluating design practices and processes for creating and management of secure software, threat modeling, and quality assurance testing using both static and dynamic analysis tools. The tutorial was conducted with stakeholders from both industry and academia with the material made available to the community at large via open-source collaboration platforms. The greater goal of his fellowship efforts is to enable community members to analyze the security of scientific software. By providing a hands-on experience through his tutorial, developers learned to provide trustworthy and secure scientific software that can mitigate threats such as, losing business and sensitive information due to a variety of potential vulnerabilities that often costs an organization thousands of dollars in patching vulnerabilities stemming from unsecured code. BBSwF provided Sukhija an opportunity to collaborate with distinguished researchers through the ECP project. These interactions helped him and his students learn more about HPC and Cybersecurity challenges, which are not reached in his department/center at the Slippery Rock University of Pennsylvania.

\paragraph*{\textbf{Amiya K. Maji}} is a Senior Computational Scientist at the Rosen Center for Advanced Computing at Purdue University. He leads the software build and test automation project targeting the university's HPC clusters, including the NSF-funded Anvil supercomputer. In this role, Amiya collaborates with faculty and researchers from various scientific domains to optimize their computational and data analysis workflows; develops training materials for the latest HPC tools and technologies; and advises scientists about writing robust software. Amiya’s research focuses on reliability and security of large-scale distributed systems. He has developed several algorithms and tools for software testing in HPC and the Internet-of-Things (IoT) domains. Amiya received the IEEE Dependability Society (DSN) Test of Time award (2022) for his doctoral research on Android platform reliability. Amiya also served as a fellow of Trusted CI (2021) where he promoted best practices for secure computing.

During his fellowship (2022), Amiya worked to simplify scientific Python package installation by streamlining environment management, dependency tracking, and runtime customizations through easy-to-use tools developed for the scientific end users. Managing Python applications is especially challenging since packages can be installed via multiple package managers, have incomplete dependencies, and there is a growing need for providing consistency across traditional batch workloads and interactive notebooks in HPC environments. He implemented the best practices for package management in the development of the \texttt{python-env-mod} tool. \texttt{python-env-mod} helps users manage their Python environments more efficiently and load runtime configurations through the familiar abstraction of environment modules. His work improved scientific productivity by automating the process of environment management to avoid user errors. The BSSwF provided Amiya the ability to network with HPC scientists who provided early feedback and suggestions about \texttt{python-env-mod}. In addition, Amiya was able to hire and mentor undergraduate students about the HPC software ecosystem.

\paragraph*{\textbf{Karan Vahi}} is a Senior Computer Scientist in the Science Automation Technologies group at the University of Southern California (USC) Information Sciences Institute. His interests and experience in the field of scientific workflows spans two decades. He is currently the lead architect and core developer of the Pegasus Workflow Management System~\cite{DEELMAN201517}. Karan's work on implementing integrity checking in Pegasus for scientific workflows won the best paper and the Phil Andrews Most Transformative Research Award at the Practice \& Experience in Advanced Research Computing (PEARC19) conference.

During Karan's fellowship (2022), he developed training materials that examine the workflow lifecycle and challenges associated with the inter-dependency of various steps such as creation, execution, monitoring and debugging in simulation and data analysis pipelines. The overall goal is to bring the use of workflows to the wider scientific community. His work walks users through the process for how to model existing simulation pipelines into workflows, how to package application code in containers, and how to execute the workflow on HPC resources and distributed computing infrastructure such as Open Science Grid. He used interactive Jupyter noteboooks to guide users on how to develop workflows that leverage application containers for running jobs using Pegasus\footnote{\url{http://pegasus.isi.edu}}. These notebooks can be used as self-guided, in classroom or virtual training material. Dissemination of the work targeting a scientific audience was done as a tutorial at PEARC 2022, and at the upcoming EScience 2022 conference. The BSSwF recognition also enabled Karan to engage with NERSC staff members on how to best expand his work to their HPC resources. 

\subsection{Honorable Mentions}

\paragraph*{\textbf{Keith Beattie}} is a Computer Systems Engineer at LBNL and holds a MS in Computer Science from San Francisco State University.  After graduation, he worked in industry for 5 years as a software and release engineer to later come back to LBNL where his career spans two decades. His interests lie upon bringing adaptive, modern, scalable and open source software engineering and development practices to academic and research contexts. He has worked on several projects bringing his expertise to LBNL for better scientific software. The BSSwF recognition (2021) has allowed him to be part of the RSE community and contribute to the discussion and plans on improving not only the state of scientific software, but also the state of careers that focus on scientific software as a primary artifact. He continues to apply the tools, processes, and rigor learned in industry, which generally has separate teams for software engineering, quality assurance, documentation, user experience, operations, and product management, revealing that the scientific community has much to learn when it comes to an understanding the commitment required to produce, and benefits gained from producing, quality scientific software. Keith considers BSSwF as one of several relatively new initiatives creating a community to address these needs.

\paragraph*{\textbf{Addi Malviya Thakur}} serves as the Group Leader of the Software Engineering Group in Computer Science and Mathematics Division's Advanced Computing Systems Research Section at Oak Ridge National Laboratory (ORNL). She is an experienced leader with a strong background in leading teams and architecting large-scale scientific software design, architecture, development. Her research interests include interconnected science and federated systems, research software engineering, operational workflows, software analytics, software frameworks, and ecosystems for science. Addi serves as the Principal Investigator of the INTERSECT-SDK program at ORNL that aims to develop a unified and coherent Software Development Kit (SDK) for federated instruments and automated experiments. The BSSwF recognition (2021) encouraged Addi to continue developing methodologies and guidelines for systematic software design and development that consider underlying hardware and data architecture and environment semantics. Her focus on design patterns for scientific software identifies a very important aspect to be improved due to the unpredictable expansion of scientific software. Her efforts benefited novel scientific applications being written as part of DOE's ECP and the ORNL's Spallation Neutron Source to adopt design patterns in their data reduction software. Her work raised awareness for code reliability, maintainability, coherency, extensibility, and usability. She hopes her effort will provide knowledge and usage guidance to thousands of people in the future and will advance the science of software development for the foreseeable future. 

\paragraph*{\textbf{Sarah E. Bratt}} is an Assistant Professor at the University of Arizona School of Information (iSchool). She obtained her PhD in Information Science \& Technology from Syracuse University in 2022. Sarah’s interests and background include developing workflows using scientific software in the context of neuroimaging and genomics data. At Syracuse University Newhouse School of Communication, she coordinated a team that employed computational notebooks including Jupyter and Orange ML to develop data management and analysis pipelines for fNIRS and EEG data. Her team created a noSQL database (Neo4j) to relate the neuro-imaging data to open source datasets. In her research, Sarah used user-experience (UX) design approaches to 1) develop a model of how genetics and genomics biologists have institutionalized data deposit; 2) assist emerging disciplines who are institutionalizing data deposit (e.g., the social and political sciences) and 3) develop HPC workflows (e.g., \cite{WorkflowBigMetadata}). The BSSwF recognition (2022) enabled Sarah to identify opportunities to shift the narrative and revise organizational structures to better support and accelerate future RSE work, such as integration in non-traditional fields. BSSwF and the ECP annual conference created an avenue for piloting a project in workflows for GenBank metadata with collaborators that support her vision to connect library and information science (LIS) research with HPC solutions. The fellowship connected her to a community of diverse scholars whose expertise and advocacy for better scientific software leadership. 

\paragraph*{\textbf{William F. Godoy}} is a Senior Computer Scientist at ORNL. He received his PhD in Mechanical Engineering from The State University of New York at Buffalo in 2009. Previous experience include a staff position at Intel Corporation and a NASA postdoctoral fellowship.
William's interests and experience are in HPC and software infrastructure for large scale scientific computing, programming models, and parallel I/O. He is a Senior IEEE Member and has served on several Computer Society venues. BSSwF (2022) gave William recognition for the RSE work performed during his career in contributing to DOE-funded HPC frameworks, aside from regular publications. BSSwF empowered him to continue fostering a community that considers better scientific software practices as essential investments. He increased his participation in related activities such as: offering tutorials at ORNL for applying modern software practices, initiatives such as the SC reproducibility, promoting publication of RSE work as an author and reviewer, and serving as a mentor in internship programs for underrepresented minorities such as the ECP Sustainable Research Pathways (SRP-HPC) and the National GEM fellowship to help build the next generation of RSEs and HPC practitioners.

\paragraph*{\textbf{Rafael M. Mudafort}} is a Senior Researcher in the National Wind Technology Center at the National Renewable Energy Laboratory (NREL). His research focuses on computational modeling of wind turbine and wind farm dynamics and controls. As a research software engineer, Rafael leads project teams focusing on software quality so that the implementation of research matches the quality of the idea itself. He is interested in the design of open source software to enable accessibility and extensibility. The BSSwF recognition (2022) has enabled Rafael to engage with the BSSw community directly, and empowered him to continue advocating for the importance of software as a primary artifact of research within NREL. He continues to advocate among DOE and NREL stakeholders to incentivize investments to improve the quality of software artifacts. His goal is to switch the default focus from the outcomes of using software to also value how the software was constructed in a manner that ensures sustainability. To achieve this goal, Rafael continues to promote software practices such as descriptions of design, architecture diagrams, documentation, performance metrics to demonstrate the value of RSE work as an important components in scientific activities.

\section{Building Synergies and Lessons Learned}
\label{sec:Synergies}

This section describes the synergies built and the lessons learned across members of the BSSwF community. These are part of the vision for BSSwF to continuously grow and promote the community beyond fellowship activities.

\paragraph*{\textbf{Awareness, recognition and networking}} are at large the best motivators for BSSwF applicants. Dan, Rafael, Addi, Sarah, and Keith dedicate much of their work to recognize software efforts among science stakeholders and build community efforts, such as US-RSE, to elevate the role of RSEs. As such, the program has created pathways between academic, national laboratory, and private sector recipients to amplify their efforts and their professional networks. 

\paragraph*{\textbf{Addressing technical challenges}}  as developing scientific software is a hard endeavour. Several recipients recognize the need to tackle difficult challenges in their proposed work such as correctness (Ignacio and Cindy), workflows (Karan and Amiya), security (Sukhija and Amiya), parallel I/O (Ritu and William), effective collaboration (Damian, Addi and Sarah). BSSwF is a way to connect and enable efforts around several technical challenges in building the next generation of scientific software.

\paragraph*{\textbf{Training}} is a key aspect in our scientific community. Much of the complexity of building scientific software is due to its dynamic nature in which new technologies and practices can result in disruptive change. In addition, gaps due to the lack of formal software development training in the broader science and engineering community is compensated with tutorials offered by those with more software expertise. BSSwF members have made training a central effort in their community involvement.

\paragraph*{\textbf{Creating venues}} that promote and growth the community for better scientific software. Cindy and Ignacio are founders and chairs of the prestigious Workshop on Software Correctness for HPC Applications running since 2017 at IEEE/ACM Supercomputing. Addi and William are part of the organization of the annual ORNL Software and Data Expo\footnote{~\url{https://softwaredataexpo.ornl.gov}}, which provides a platform for stakeholders to celebrate RSE efforts in a national laboratory environment. Dan has spent much of his efforts in building the US-RSE association to provide a sense of community around software independently of particular scientific domains it supports.

\paragraph*{\textbf{Representation}} is a key aspect to advance the community. Damian, Sarah, Addi and William have centered their efforts around increasing participation of underrepresented sectors of the population as well as incorporating other software stakeholders that are not always part of the conversation.

\subsection{Lessons Learned}
A few lessons have arisen from the experiences of the BSSwF community since established in 2018. One important lesson is the increase of the fellowship amount from \$10,000 to \$25,000 to attract more applicants seeking to make an impact in their proposed work. The takeaway is that the award amount could be revised as needed.  Another important lesson is the dissemination and awareness of efforts around HPC work that only formally engages with other research communities that could benefit from it.  This disconnect can be addressed through initiatives like BSSwF and by empowering RSEs to be ambassadors and a proxy for closing these gaps.

The gender and diverse representation of the BSSwF community provides a great opportunity to develop mentorship relationships to strengthen the retention and advancement of the diverse networks around the community. Initiatives like BSSwF are uniquely positioned to catalyze programs to address the representation imbalance in software engineering industry and academia through its initiatives, e.g., programming, workshops, and scholarships. 

\section{Conclusions}
\label{sec:Conclusions}
Scientific software is crucial and more complex than ever. The end of Moore's Law, the avalanche of cloud services, the heterogeneity of computing hardware, the growth of the RSE movement, and the data tsunami from AI/ML workflows as major drivers of modern computing are disruptive to traditional software design paradigms, emphasizing the need to foreground issues of how we build future software. From the experiences presented in this paper we believe that BSSwF is a strong initiative that recognizes and elevates the role of those researching, developing, and maintaining key software components in the critical path of the scientific enterprise. The effort to bringing a diverse pool of individuals has been tremendously successful in building next-generation scientific software to accelerate innovation and discovery. Our shared view is that the future of RSEs must incorporate initiatives like BSSwF that value and foster these efforts and create the required synergies among the scientific communities to achieve their goals.

\section{Acknowledgments}
David E. Bernholdt thanks the other members of the Better Scientific Software Fellowship Executive Committee for the honor of representing them in this publication: Varun Chandola (National Science Foundation), Mike Heroux (Sandia National Laboratories), Lois Curfman McInnes (Argonne National Laboratory), Hai Ah Nam (Lawrence Berkeley National Laboratory), and Shelly Olsan (Krell Institute).
William F. Godoy would like to acknowledge Arthur Barney Maccabe and Jay Jay Billings for being early enablers of the Software Expo for the RSE community at Oak Ridge National Laboratory.
This work was supported by the Better Scientific Software Fellowship Program, funded by the Exascale Computing Project (17-SC-20-SC), a collaborative effort of the U.S. Department of Energy (DOE) Office of Science and the National Nuclear Security Administration; and by the National Science Foundation (NSF) under Grant No. 2154495.

\bibliographystyle{IEEEtran}
\bibliography{IEEEabrv,paper.bib}

\end{document}